\documentclass[submission, Phys]{SciPost}
\usepackage{graphicx,rotating,amsmath,amsfonts,amssymb,delarray,color}
\usepackage{dsfont}
\usepackage[T1]{fontenc}
\usepackage{multirow}
\usepackage{comment}
\usepackage{physics}
\usepackage{hyperref}
\hypersetup{
     colorlinks = true,
     linkcolor = blue,
     anchorcolor = blue,
     citecolor = blue,
     filecolor = blue,
     urlcolor = blue
     }

\newcommand{\e}{\text{e}}
\newcommand{\im}{\text{i}}

\def\12{\frac{1}{2}} 

\usepackage{hyperref}

\begin{document}

\begin{center}{\Large \textbf{Bounds on the entanglement entropy by the number entropy\\ in non-interacting fermionic systems
}}\end{center}

\begin{center}
M. Kiefer-Emmanouilidis\textsuperscript{1,2},
R. Unanyan\textsuperscript{1},
J. Sirker \textsuperscript{2},
M. Fleischhauer \textsuperscript{1}
\end{center}

\begin{center}
{\bf 1} Department of Physics and Research Center OPTIMAS, University of Kaiserslautern, 67663 Kaiserslautern, Germany
\\
{\bf 2} Department of Physics and Astronomy, University of Manitoba, Winnipeg R3T 2N2, Canada
\\
\end{center}

\begin{center}
\today
\end{center}

\section*{Abstract}
Entanglement in a pure state of a many-body system can be
characterized by the R\'enyi entropies
$S^{(\alpha)}=\ln\textrm{tr}(\rho^\alpha)/(1-\alpha)$ of the reduced
density matrix $\rho$ of a subsystem. These entropies are, however,
difficult to access experimentally and can typically be determined for
small systems only. Here we show that for free fermionic systems in a Gaussian state and with
particle number conservation, $\ln S^{(2)}$ can be tightly bound by
the much easier accessible R\'enyi number entropy $S^{(2)}_N=-\ln \sum_n
p^2(n)$ which is a function of the probability distribution $p(n)$ of
the total particle number in the considered subsystem only. A
dynamical growth in entanglement, in particular, is therefore always
accompanied by a growth---albeit logarithmically slower---of the
number entropy. We illustrate this relation by presenting numerical
results for quenches in non-interacting one-dimensional lattice models
including disorder-free, Anderson-localized, and critical systems with
off-diagonal disorder.

\vspace{10pt}
\noindent\rule{\textwidth}{1pt}
\tableofcontents\thispagestyle{fancy}
\noindent\rule{\textwidth}{1pt}
\vspace{10pt}

\section{Introduction}
\label{Intro}
The entanglement between two parts of a many-body system in a pure
state can be characterized by the R\'enyi entropies
$S^{(\alpha)}=\ln\textrm{tr}(\rho^\alpha)/(1-\alpha)$. The
von-Neumann entanglement entropy is given by
$S=S^{(1)}\equiv\lim_{\alpha\to 1}S^{(\alpha)} = -\textrm{tr}(\rho\ln \rho)$. Their
time evolution, for example following a quantum quench, offers important
insights into the dynamics of the many-body system.  Except for very
small systems, where the reduced density matrix $\rho$ can be obtained
from quantum-state tomography \cite{Blatt1}, these entropies are
difficult to access experimentally. For systems with particle number
conservation, the R\'enyi entropies can be expressed as
\begin{equation}
S^{(\alpha)}=\frac{1}{1-\alpha}\ln\left(\sum_n p^\alpha(n)\,\mbox{tr}\rho^\alpha(n)\right) = S^{(\alpha)}_N+S^{(\alpha)}_{\rm conf}\, .
\end{equation}
Here $p(n)$ is the probability distribution of the particle number $n$
in the partition and $\rho(n)$ is the block of the reduced density
matrix with fixed particle number $n$, normalized such that
$\mbox{tr}\rho(n) = 1$. The R\'enyi number entropy
$S^{(\alpha)}_N=\ln[\sum_n p^\alpha(n)]/(1-\alpha)$ is then the part
of the entanglement due to number fluctuations only (i.e., in a system
where only one configuration for each possible particle number $n$
exists we would have $S^{(\alpha)}=S^{(\alpha)}_N$) and
$S^{(\alpha)}_{\rm conf}$ describes the additional entanglement due to
the existence of several configurations for a given $n$. This
configurational entropy takes the particularly simple form,
$S_{\rm{conf}} =-\sum_n p(n)\,\mbox{tr}[\rho(n)\ln\rho(n)]$, in the limit
$\alpha\to 1$ \cite{Klich-arxiv2006}. The corresponding number entropy
$S_N$ has been measured very recently in an experiment on a cold
atomic gas
\cite{Lukin-Science2019}. The source of the number entropy are fluctuations 
induced by particle transport. $S_{\rm conf}$, on the other hand, is
determined by the full microscopic counting statistics. $S_N$ is thus
much easier measurable in experiments and can also be accessed
theoretically using conformal field theory (CFT)
\cite{Calabrese2019}.

Here we prove that for non-interacting fermions on a lattice where the
particle number is conserved, the second R\' enyi entropy can be bounded
from above and below by 
\begin{equation}
\frac{1}{e\pi}\exp(2S_N^{(2)})-\frac{1}{6}\,\lesssim\, S^{(2)} \,\lesssim \, \frac{\ln 2}{\pi}\exp(2S_N^{(2)}),
\label{eq:result}
\end{equation}
where $S_N^{(2)}$ is the second R\'{e}nyi number
entropy. Thus the size and time dependence of the entanglement is
directly linked to that of the number entropy. A dynamical growth of
entanglement in any non-interacting fermion system, in particular,
implies that the number entropy grows as well, albeit logarithmically
slower. Vice versa, in a fully localized phase, where the fluctuations
of the particle number of a partition are expected to saturate, the
number of accessible configurations and thus entanglement can no
longer increase either.

Our paper is organized as follows. In Sec.~\ref{Bounds} we present the
proof for the lower and upper bounds in Eq.~\eqref{eq:result}. In
Sec.~\ref{Numerics} we exemplify the usefulness of these bounds based
on numerical data for the time evolution of the entropies after
quantum quenches in one-dimensional fermionic lattice models with and
without disorder. This includes, in particular, the interesting case
of off-diagonal disorder (bond disorder), where the von Neumann
entropy shows a very slow $\ln\ln t$ increase in time
\cite{IgloiSzatmari,Zhao16,Vosk2014,Vosk2013}, while the number entropy
scales as $\ln\ln\ln t$. Extensions to fermionic models with
interactions will be discussed elsewhere \cite{Kiefer-2020-2}.

\section{Bounds on the R\'enyi entropy by the number entropy}
\label{Bounds}
In the following, we will establish a relation between the second
R\'{e}nyi entropy $S^{(2)}= - \ln \textrm{tr}(\rho^2)$ of a quantum
state $\rho$, which we will refer to as purity entropy, and the
corresponding number entropy $S_N^{(2)} = -\ln \sum_n
p_n^2$. Specificially, we consider models of non-interacting fermions
with particle number conservation.

\subsection{Lower bound on the purity entropy}
Since the number entropy does not account for the different
configurations of particles in the considered subsystem, a trivial
lower bound for the purity entropy is given by
\begin{equation}
S^{(2)} \ge S^{(2)}_N \, .
\label{lower_bound_1} 
\end{equation}
This is, however, in most cases only a very weak bound. An alternative
and often much better lower bound can be obtained using the relation
between $S^{(2)}$ and the particle number fluctuations $\Delta n^2$
derived in \cite{Muth-PRL2011}
\begin{equation}
4\ln(2)\Delta n^{2}\geq S^{\left(  2\right)  }\geq2\Delta n^{2}.
\label{Estimations_Reni_second}
\end{equation}
%
From the right hand side of Eq.~(\ref{Estimations_Reni_second})
together with the modified version of Shannon's inequality for
discrete variables \cite{Cover}
\begin{equation}
\frac{1}{2}\ln\left[  2\pi e\left(
\Delta n^{2}+\frac{1}{12}\right)  \right]  \ge S_N \ge S_N^{(2)},
\label{Cover}
\end{equation}
we find the alternative lower bound for the purity entropy  
\begin{equation}
S^{\left(  2\right)  }\geq\frac{1}{e\pi}\exp\left(  2S_{N} \right)  -\frac{1}{6}\ge \frac{1}{e\pi}\exp\left(  2S_{N}^{\left(
2\right)  }\right)  -\frac{1}{6}\, .\label{lower_bound_2}%
\end{equation}
For $S_N^{(2)} > \ln (e \pi/6) /2 \approx 0.18 $ this bound is
positive and for $S_N^{(2)} \gtrsim 1.25$ it is a stricter lower bound
than the trivial relation (\ref{lower_bound_1}).\\


\subsection{Upper bound on the purity entropy}
In order to derive an upper bound for the purity entropy we make use
of the fact that the quantum state $\rho$ for a non-interacting
fermionic system in any dimension is completely determined by its
single-particle correlations and has a Gaussian form.  This applies in
particular to all eigenstates of free-fermion Hamiltonians and to all
time-evolved states under such Hamiltonians if the initial state is
Gaussian.  Since we assume, furthermore, total particle number
conservation, $\rho$ can be represented as
\cite{Chung2001,Peschel2004,Peschel2009}
\begin{equation}
\rho=\frac{1}{Z}\exp\left(-
\sum\limits_{mn}
c_{m}^{\dagger}C_{mn}c_{n}\right), \label{Density_matrix}%
\end{equation}
where ${c}_m ({c}^\dagger_m)$ are the fermionic annihilation
(creation) operators at lattice site $m$.
Here $\mathbf{C}$ is a Hermitian matrix which is determined entirely
by the matrix $\mathbf{f}$ of (normal) single-particle correlations
\begin{equation}
f_{mn}=\left\langle c_{m}^{\dagger}c_{n}\right\rangle =\textrm{tr}\left(  \rho\, 
c_{m}^{\dagger}c_{n}\right)  =\left[  \frac{1}{1+e^\mathbf{C}}\right]  _{mn},\label{correlations}
\end{equation}
and 
$Z=\textrm{tr}\left\{\exp\bigl(\, -\sum_{nm}c^\dagger_n C_{nm} c_m\bigr)\right\}$.

We will now show that the particle number fluctuations $\Delta n^2$ in
a partition are bounded from above by the R\'{e}nyi number entropy
$S_N^{(2)}$. Making use again of relation
(\ref{Estimations_Reni_second}) this will then result in an upper
bound on the purity entropy in terms of $S_N^{(2)}$. To do so, it is
useful to introduce the moment generating function of the total
particle number $\hat N$ in the partition \cite{Klich2} 
\begin{equation}
\chi(\theta) \equiv \left\langle e^{\im\theta \hat N}\right\rangle = \textrm{tr} \Bigl\{ \rho \, e^{\im\theta \hat N}\Bigr\} = \sum_n p(n)\, e^{\im n\theta}.
\end{equation}
For Gaussian fermionic states, the generating function can be written as a determinant
\cite{Schoenhammer-PRB-2007}
\begin{equation}
\chi(\theta) = \det \Bigl[\mathbf{1} + (e^{\im\theta} -1) \frac{\mathbf{1}}{\mathbf{1}+e^\mathbf{C}}\Bigr].\nonumber
\end{equation}
Making use of Parsevals theorem one then finds 
\begin{eqnarray}
\sum_n p_n^2 &=& \frac{1}{2\pi}\int_0^{2\pi}\!\! \! d\theta\, \, \bigl\vert \chi(\theta)\bigr\vert^2 =  \frac{1}{2\pi}\int_0^{2\pi}\!\! d\theta\, \Bigl\vert \det \Bigl[\mathbf{1} + (e^{\im\theta} -1) \frac{\mathbf{1}}{\mathbf{1}+e^C}\Bigr]\Bigr\vert^2\label{purity} \\
&=& \frac{1}{2\pi} \int_0^{2\pi}\!\! d\theta\, \frac{\det\left( \mathbf{1}+2e^\mathbf{C}\cos\theta+e^{2\mathbf{C}}\right)  }{\det^{2}\left(
\mathbf{1}+e^\mathbf{C}\right)  } = \frac{1}{2\pi} \int_0^{2\pi}\!\! d\theta\, \det\left(\mathbf{1}-\mathbf{G} +\mathbf{G} \cos\theta\right). \nonumber
\end{eqnarray}
In the last equation we introduced the matrix 
\begin{equation}
\mathbf{G}=\frac{2 e^\mathbf{C}}{\left(\mathbf{1}+e^\mathbf{C}\right)^2} = 2\, \mathbf{f}(\mathbf{1}-\mathbf{f}) \le \frac{1}{2} \mathbf{1}.\label{G}
\end{equation}
We see that the argument in the last line of Eq.~(\ref{purity}) is a
positive-definite matrix.  Thus we can apply the arithmetic-geometric
inequality to get an upper bound on $\det\left(\mathbf{1}-\mathbf{G}
+\mathbf{G} \cos\theta\right)$. Denoting the lattice size as $M$ we
find
\begin{equation}
\sum_n p_n^2 \le \frac{1}{2\pi} \int_0^{2\pi}\!\! d\theta\,  \left[1 + \frac{(\cos\theta - 1) \textrm{tr}(\mathbf{G})}{M}\right]^M \rightarrow \frac{1}{2\pi} \int_0^{2\pi}\!\! d\theta\, \exp\Bigl[(\cos\theta -1) \textrm{tr}(\mathbf{G})\Bigr],
\end{equation}
where the second line holds in the thermodynamic limit $M\to\infty$. The integral can be calculated elementary in terms of
the modified Bessel function of the first kind $I_0(x)$ resulting in
\begin{equation}
\sum_n p_n^2 \le \exp\bigl(-\textrm{tr}(\mathbf{G})\bigr)\, I_0\bigl(\textrm{tr}(\mathbf{G})\bigr).
\label{purity-2}
\end{equation}
Furthermore, we see from Eq.~(\ref{G}) that the trace of the matrix
$\mathbf{G}$ gives the fluctuations of the total particle number
\begin{equation}
\textrm{tr}(\mathbf{G})= 2\, \tr\bigl(\mathbf{f}(\mathbf{1}-\mathbf{f})\bigr) = 2\, \Delta n^2  \, .
\end{equation}
Combined with Eq.~(\ref{purity-2}) we therefore find for the number entropy
\begin{eqnarray}
S_N^{(2)} = -\ln \sum_n p_n^2 \ge 2 \Delta n^2 - \ln\Bigl(I_0\bigl(2 \Delta n^2\bigr)\Bigl).
\label{bound_IO}
\end{eqnarray}
Using the asymptotic expansion of the modified Bessel function in the
limit of large $\Delta n^2$, this expression can be simplified to
\begin{equation}
S_N^{(2)}\ge  \frac{1}{2}\ln (4\pi \Delta n^2).
\end{equation}
In the opposite limit of small $\Delta n^2$ the contribution of the
modified Bessel function in Eq.~(\ref{bound_IO}) can be neglected and
we find instead
\begin{equation}
S_N^{(2)}\ge  2\Delta n^2.
\end{equation}
Now making use of the left hand side of the inequality in
Eq.~(\ref{Estimations_Reni_second}), we eventually arrive at an upper
bound on the purity entropy in terms of the R\'{e}nyi number
entropy. This bound can be written explicitly in the two limiting
cases of either small values of $\Delta n^2$
\begin{equation}
S^{(2)} \lesssim (2\ln2)\, S_N^{(2)},\label{upper_bound1}
\end{equation}
or large values of $\Delta n^2$
\begin{equation}
S^{(2)} \lesssim \frac{\ln 2}{\pi} \exp\Bigl( 2 S_N^{(2)}\Bigr).
\label{upper_bound}
\end{equation}
Eqs.~(\ref{upper_bound1},\ref{upper_bound}) and (\ref{lower_bound_2})
are the main results of our paper. They show that the entanglement
quantified by the logarithm of the purity entropy $S^{(2)}$
is bounded both from below \textit{and} above by the number
entropy. As a consequence, a growth of entanglement in free fermionic
systems is always accompanied by a logarithmically slower growth of
the number entropy.

\section{Time evolution of entropies in free fermionic systems}
\label{Numerics}
Next, we illustrate our results by considering one-dimensional
tight-binding models of non-interacting fermions. By applying a
Jordan-Wigner transformation, these models can alternatively also be
seen as spin-$1/2$ XX chains. We discuss free fermions without
disorder in Sec.~\ref{example1}, with potential disorder leading to
Anderson localization in Sec.~\ref{example2}, and with bond
(off-diagonal) disorder resulting in a critical system in
Sec.~\ref{example3}.

The Hamiltonian for all these systems has the same structure
\begin{equation}
\label{Ham}
{H}= \sum_{j=1}^{L-1} J_j (\hat{c}_j^\dagger\hat{c}_{j+1} +h.c.) +\sum_{j=1}^L D_j\hat{c}_j^\dagger\hat{c}_{j},
\end{equation}
with hopping amplitudes $J_j$, onsite potentials $D_j$, system size
$L$, and open boundary conditions. In the following, we always
consider the R\'enyi and number entropies for a partition of size
$l=L/2$. We are interested in the time evolution of the entanglement
entropies following a quantum quench starting from an initial state
$\ket{\Psi_\mathrm{ini}}$ which is not an eigenstate of the
Hamiltonian \eqref{Ham}.

We use exact diagonalization (ED) methods to obtain the eigenvalues of
the reduced density matrix $\rho(t)$, see Eq.~(\ref{Density_matrix}),
which is calculated from the time-evolved state $\ket{\Psi (t)} =
\e^{-i\hat{H}t} \ket{\Psi_\mathrm{ini}}$ following
Refs.~\cite{Chung2001,Peschel2004,Peschel2009}. From the eigenvalues
$f_m$ of the correlation matrix $\mathbf{f}$ in
Eq.~(\ref{correlations}), we can directly obtain all quantities of
interest efficiently. This includes the von-Neumann and purity entropies
\begin{align}
\begin{split}
S &= -\sum_mf_m \ln(f_m)-\sum_m(1-f_m)\ln(1-f_m), \\
S^{(2)} &= -\sum_m \ln(1-2f_m(1-f_m)), 
\end{split}
\end{align}
as well as the number distribution which can be calculated from the corresponding characteristic function 
\begin{equation}
p(n) = \frac{1}{l+1}\sum_{k=0}^{l}\exp(-\im\frac{2\pi k n}{l+1}) \chi(k),\quad \chi(k) = \ev{\mathrm{e}^{\im\frac{2\pi k}{l+1} \hat{N}}} = \prod_m\left(1+\left(\e ^{\im\frac{2\pi k}{l+1}} -1 \right) f_m \right).
\end{equation}

\subsection{Free fermions without disorder}
\label{example1}
The case of fermions on a one-dimensional lattice has been studied
extensively in the past, both analytically using conformal field
theory \cite{Calabrese2009} as well as numerically, see
e.g.~Ref.~\cite{Zhao16}. The conformal field theory results show that
the von-Neumann entropy as well as all R\'{e}nyi entropies increase
linearly in time in the thermodynamic limit although with a slope
which is non-universal. Here we choose a density-wave state with a
fermion on every second site,
\begin{equation}
\ket{\Psi_\mathrm{ini}}=
\prod^l_{j=1}\hat{c}_{2j}^\dagger \ket{0},
\label{Neel}
\end{equation}
as initial state. Note, however, that the results are qualitatively
the same for any generic initial product state. 
%
\begin{figure}
\includegraphics[width=1\textwidth]{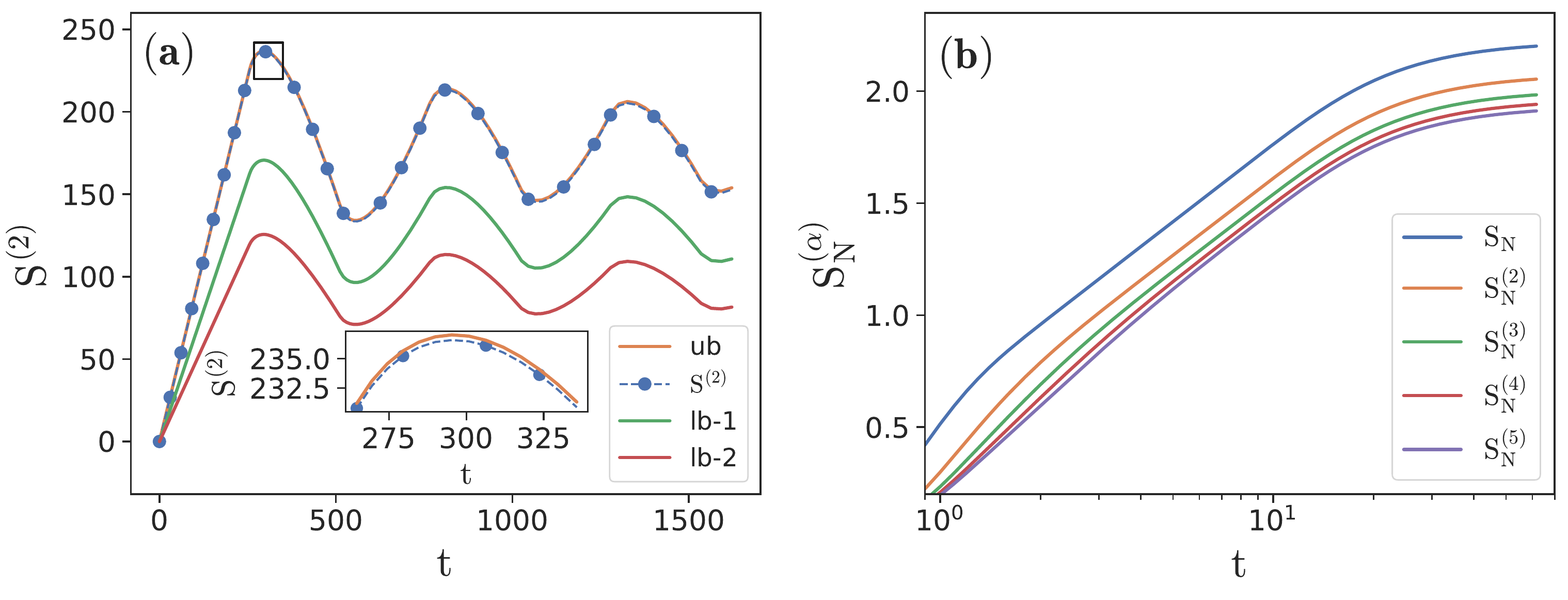}
\caption{(a) $S^{(2)}(t)$ for a quench from the initial state \eqref{Neel} using the Hamiltonian \eqref{Ham} 
with $J_j=1$, $D_j=0$, and system size $L=1024$. The purity entropy $S^{(2)}$ grows linearly until it reaches its maximum at $t\approx
l/v$. The upper bound (ub), Eq.~(\ref{upper_bound}), and the two lower bounds
from Eq.~(\ref{lower_bound_2}), (lb-1) $\exp(2 S_N)/(e\pi)-1/6$, and (lb-2) $\exp(2 S_N^{(2)})/(e\pi)-1/6$,
encapsule $S^{(2)}$. Note that the
upper bound is tight, see inset.
(b) R\'enyi number entropies $S_N^{(\alpha)}(t)$ for $20$ 
Gaussian waves who are initially spaced at equal distances and whose
width increases linearly in time, see Eq.~\eqref{Gauss},  for $\nu=1$. We find
$S_N^{(\alpha)}(t)=\mbox{const}+\frac{1}{2}\ln t$. The saturation at
long times is a finite-size effect.}
\label{tight_binding_Fig_gaussian}
\end{figure}
%
The numerical results in Fig.~\ref{tight_binding_Fig_gaussian}(a) show that
$S^{(2)}(t)$ increases linearly in time until the particle-hole pairs
created by the quench reach the boundaries of the partition of size
$l=L/2$. This happens for times $t\sim l/v\approx l/2$
\cite{Zhao16,Calabrese2009} where $v\approx 2$ is the velocity of the
excitations. For times $t>l/v$ boundary effects dominate the dynamics
and $S^{(2)}$ is oscillating around an average value which depends on
the size of the partition. Fig.~\ref{tight_binding_Fig_gaussian}(a) confirms that the
lower and upper bounds obtained here are valid for all times,
including long times where boundary effects dominate. The upper bound
\eqref{upper_bound}, in particular, is a very tight bound for all times 
in this case, see the inset of Fig.~\ref{tight_binding_Fig_gaussian}(a).

Based on the bounds and verified by the numerical results above we
find that the R\'{e}nyi number entropy grows as $S^{(2)}_N(t) \sim
\ln t$ for free fermions on a one-dimensional lattice without
disorder. This logarithmic growth can be understood as follows:
Consider a quench in a half-filled system where at long times each
arrangement of particles has approximately the same probability. Then
for a system of size $2L$ we have $L$ particles. If we cut the system
in two halfs of size $L$, the probability to find $k$ particles in
one half is given by
\begin{equation}
p(k,L) = \frac{{{L}\choose{k}} {{L}\choose{L-k}}}{{{2L}\choose{L}}} \, .
\end{equation}
For $1\ll k < L$ we can approximate this distribution by a normal distribution
\begin{equation}
\tilde p(k,L) = \frac{2}{\sqrt{L\pi}}\exp\left[-\frac{4}{L}\left(k-\frac{L}{2}\right)^2\right] \, .
\end{equation}
We can now obtain the R\'enyi number entropies by integrating over the
continuous distribution
\begin{equation}
\label{NumberSimple}
S_N^{(\alpha)}(L) \approx \frac{1}{1-\alpha}\ln\left(\int_{-\infty}^\infty dk\,\tilde p^{\,\alpha}(k,L)\right) = \ln\left[\frac{\sqrt{\pi}}{2\alpha^{1/(2-2\alpha)}}\right]+\frac{1}{2}\ln L\, . 
\end{equation}
The von-Neumann number entropy can be obtained by 
\begin{equation}
S_N(L)=\lim_{\alpha\to 1}
S^{(\alpha)}_N(L)
\approx \frac{1}{2} \left(1+\ln\left[\frac{L \pi }{4}\right]\right) \, .
\end{equation}
If we now consider excitations which spread ballistically $\sim vt$
then we have regions of size $2L\sim vt$ in which each arrangement of
particles has approximately equal probability. Putting this into the
results for the number entropy and the R\'enyi number entropies we
obtain the final result
\begin{equation}
\label{main}
S_N^{(\alpha)}(t) = \mbox{const} + \frac{1}{2}\ln t 
\end{equation}
which includes the von-Neumann case ($\alpha\to 1$). Note that the
constants do depend on the microscopic details of the model but are
monotonically decreasing with $\alpha$, see
Eq.~\eqref{NumberSimple}. Thus we conclude that all R\'enyi number
entropies behave the same qualitatively and
$S_N^{(\alpha)}>S_N^{(\alpha+1)}$.

An alternative perspective to understand the logarithmic
spreading---more closely related to the numerical simulations--- can
be obtained by considering Gaussian waves
\begin{equation}
\label{Gauss}
|\Psi_i(x,t)|^2 = \frac{1}{\sqrt{4\pi \nu t^{2}}}\exp\left(-\frac{(x-x_i)^2}{4 \nu t^{2}}\right)
\end{equation}
with initial positions $x_i$ spread evenly along a line. Here $\nu$ is a
constant and the width of the Gaussian wave is increasing linearly in
time. The probability to find the particle $i$ at $x>0$ is then given
by
\begin{equation}
P(x_i,t)=\int_0^\infty |\Psi_i(x,t)|^2 = \frac{1}{2}\left(1+\mbox{erf}\left(\frac{x_i}{2\sqrt{\nu t^{2}}}\right)\right) \, .
\end{equation}
If we have $N$ particles in total then the probability to find $k$ at $x>0$ is
\begin{equation}
p(k,t) = \!\!{\sum_{n_i \in \{0,1\}}\!\!}^\prime
\;\;\prod_{i=1}^N [P(x_i,t)]^{n_i}[1-P(x_i,t)]^{1-n_i}
\end{equation}
The sum $\sum^\prime$ is over all permutations of the $\{n_i\}$ and has to be evaluated with the
constraint $\sum_{i=1}^N n_i =k$. It can be
directly evaluated if $N$ is not too large. Results for $N=20$
particles are shown in Fig.~\ref{tight_binding_Fig_gaussian} (b) and confirm
Eq.~\eqref{main}.

\subsection{Anderson Localization}
\label{example2}
Static potential disorder in an isolated quantum system of
non-interacting particles can induce Anderson localization (AL),
defined as the absence of particle diffusion
\cite{Anderson1958}. For one and two dimensions, Anderson localization 
occurs for any strength of disorder $D$ \cite{Abrahams1979}. For a
one-dimensional system we can extract the localization length $\xi$ in
dependence of energy $\epsilon$ and disorder strength $D$ using a
transfer matrix approach as described in \cite{Mueller2016}. If we
quench a one-dimensional system with potential disorder we thus expect
that for times $t\gg \xi/v$ both the number and configurational
entropies will stop increasing. 

\begin{figure}[htb]
\includegraphics[width=1\textwidth]{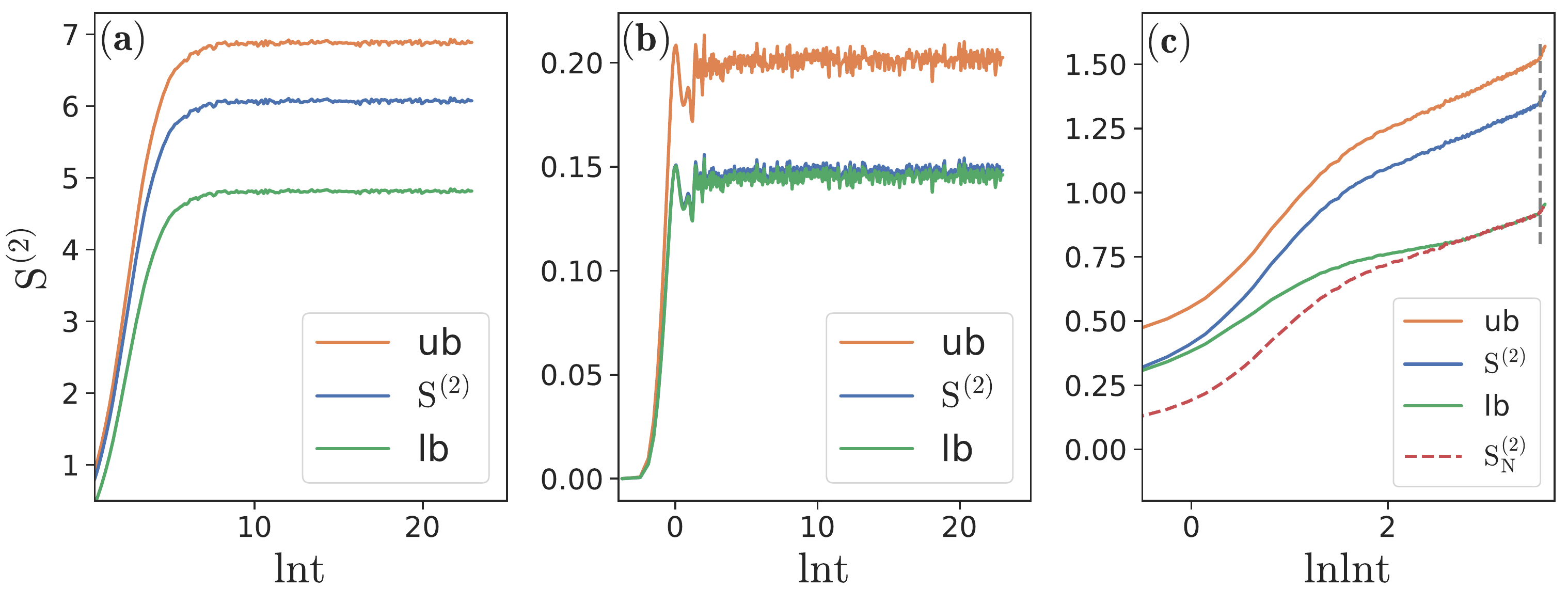}
\caption{(a) In the weakly disordered case $D=2$, the upper bound (ub), $(\ln 2/\pi)
\exp(2 S_N^{(2)})$ is 
quite tight for the regime in which $S^{(2)}(t)$ grows.
The lower bound (lb) shown is $\exp(2 S_N)/(e\pi)-1/6$.
 (b) For strong
disorder, $D=20$, the entanglement remains very small and
Eq.~(\ref{upper_bound1}), $2\mathrm{ln}(2)S^{(2)}_\mathrm{N} $, is the better upper bound (ub). For the same reason $S_N^{(2)}$ (lb)
is a better lower bound. In both
cases averages over 2000 disorder realizations and initial states are
shown.
(c) $S^{(2)}(t)$ for bond disorder after a quench starting from the half-filled 
random product state \eqref{psi_random} for $L=1024$ sites and $20000$
disorder realizations. (ub) corresponds to $(\ln2/\pi)\, \exp(2S_N^{(2)})$, (lb) to the maximum of $S_N^{(2)}$ and
$\exp(2 S_N)/(\pi e) -1/6$.
We confirm that $S^{(2)}\sim
\ln\ln t$. At long times, $S^{(2)}$ and the upper bound (ub) \eqref{upper_bound} based on the number entropy 
only differ by a constant shift. The grey dotted line signals the
point in time where double precision is no longer sufficient to obtain
reliable results, see also Ref.~\cite{Zhao16}.}
\label{fig_anderson_bond_dis}
\end{figure}

To study the Anderson case numerically, we set the hopping in
Eq.~\eqref{Ham} uniformly to $J_j=1$ and draw random values for the
potential from a box distribution $D_j\in [-D/2,D/2]$. We now quench
the system from initial random product states at half filling
\begin{equation}
\ket{\Psi_\mathrm{ini}} = \prod_{j=1}^l \eta_j\hat{c}_{j}^\dagger \ket{0},
\label{psi_random}
\end{equation}
where $\eta_j\in \{0,1\}$ is random, and half-filling is imposed by
requiring $\sum_j \eta_j=l/2$. The random product state will on
average yield a state with energy $\epsilon =0$ and we consider both a
weakly disordered case, $D=2$, and a strongly disordered case,
$D=20$. Since we consider systems large compared to the localization
length we do not expect a qualitative difference in the time
dependence of the purity entropies in the two cases: there will be an
increase of entropy in time until a constant value is reached that
does depend on the localization length $\xi$ but is independent of
system size $L$ if $L\gg\xi$. Our numerical simulations, shown in
Fig.~\ref{fig_anderson_bond_dis}(a-b), verify this
behavior. Furthermore, they also verify the lower and upper bounds in
terms of the number entropy.  Note that $S^{(2)}$ is bounded quite
tightly from below by $S^{(2)}_N$ for small values of $S^{(2)}$, see
Fig.~\ref{fig_anderson_bond_dis}(b). Here the trivial lower bound,
Eq.~(\ref{lower_bound_1}), is the better bound since both the purity
entropy and the corresponding number entropy are quite small in the
localized regime.
%

\subsection{Bond disorder}
\label{example3}
As the third example, we consider the Hamiltonian \eqref{Ham} with
bond disorder, $J_j\in (0,1]$. It is known that this system in the
thermodynamic limit is at an infinite randomness fixed point
\cite{Eggarter1978,Fisher1994,Fisher1992,Fisher1995}. The mean localization 
length scales as $\xi_\mathrm{loc}(\epsilon ) \sim | \ln(\epsilon
)|^\Psi $ with $\Psi$ being the critical exponent. The system
therefore shows a localization-delocalization transition as a function
of energy for $\epsilon\to 0$. The entanglement dynamics of this model
has been investigated previously in Ref.~\cite{Zhao16} and of the
related transverse Ising chain in Ref.~\cite{IgloiSzatmari}.

Starting from the random half-filled state in Eq.~(\ref{psi_random})
we show in Fig.~\ref{fig_anderson_bond_dis}(c) the time evolution of
$S^{(2)}(t)$ obtained from exact diagonalizations of a system with
$L=1024$ lattice sites over very long times. One notices an extremely
slow, but monotonic double-logarithmic increase of $S^{(2)}$ in time
consistent with the results in Ref.~\cite{Zhao16}. The data provide,
furthermore, verification of the upper and lower bounds for $S^{(2)}$
in terms of the number entropy derived in Sec.~\ref{Bounds}. We
conclude that in the critical bond disordered case the number entropy
in the thermodynamic limit also grows without bounds but extremely
slowly, $S^{(2)}_N\sim \ln \ln\ln t$.
%
\section{Conclusions}
In this paper we have considered the entanglement properties of Gaussian states of
non-interacting fermions with particle number conservation. We have
proven that for any such system---with and without disorder, on
arbitrary lattice geometries, and in arbitrary dimensions---the second
R\'enyi entanglement entropy $S^{(2)}$ can be bounded from {\it above
and below} by the corresponding number entropy $S_N^{(2)}$. Our result
implies an asymptotic scaling $S^{(2)}\propto\exp(S_N^{(2)})$, i.e., a
growth in the entanglement entropy always implies a growth, albeit
logarithmically slower, of the number entropy and vice versa. While
the precise upper and lower bounds have been derived for $S^{(2)}$,
all R\'enyi entropies are expected to show the same asymptotic scaling
with time or length. The connection between a growth in the
entanglement entropy and a logarithmic slower growth in the
corresponding number entropy is thus expected to hold for all R\'enyi
entanglement entropies including the von-Neumann entanglement entropy.

Apart from being of fundamental importance for our understanding of
entanglement in fermionic systems with particle number conservation,
the bounds derived here are also useful for experiments on cold atomic
gases. In such systems a measurement of the particle-number
distribution function $p(n,t)$ is possible \cite{Lukin-Science2019}
allowing to obtain any R\'enyi number entropy. Determining the entire
configurational entropy and thus the full entanglement entropy
experimentally, on the other hand, remains an open issue. Here our
results provide an avenue to obtain the asymptotic scaling of the
entanglement entropy from $p(n,t)$ alone.

An interesting  question is if similar relations between
entanglement and number entropies also exist for interacting fermionic
systems with particle number conservation. This question will be
studied in a forthcoming publication \cite{Kiefer-2020-2}.

\section*{Acknowledgements}
J.S. acknowledges support by the Natural Sciences and Engineering
Research Council (NSERC, Canada) and by the Deutsche
Forschungsgemeinschaft (DFG) via Research Unit FOR 2316. We are
grateful for the computing resources and support provided by Compute
Canada and Westgrid. M.K., R.U., and M.F. acknowledge financial
support from the Deutsche Forschungsgemeinschaft (DFG) via SFB TR 185,
project number 277625399. M. K. would like to thank J. L\'eonard and
M. Greiner for hospitality and fruitful discussions.

\end{document}